\documentclass[preprint,aps,amsmath,amssymb,12pt]{revtex4}

\usepackage{epsfig}
\usepackage{slashed}
\usepackage{graphicx}
\usepackage{multirow,color}
\usepackage{amsmath}
\usepackage{float}
\usepackage{diagbox}
\usepackage{CJK}
\usepackage{color}
\usepackage{xcolor}
\usepackage{times}
\usepackage{subfigure}
\usepackage{bm}
\usepackage{braket}
\usepackage{booktabs}
\usepackage{array}
\usepackage[mathscr]{euscript}
\usepackage{caption}
\usepackage{makecell}

\makeatletter

\newcommand{\Rmnum}[1]{\expandafter\@slowromancap\romannumeral #1@}
\makeatother
\graphicspath{{fig/}}
\begin{document}
\title{Detecting the coupling of axion-like particles with fermions at the ILC}
\author{Chong-Xing Yue$^{1,2}$}
\thanks{cxyue@lnnu.edu.cn}
\author{Han Wang$^{1,2}$}
\thanks{wangwanghan1106@163.com (corresponding author)}
\author{Yue-Qi Wang$^{1,2}$}
\thanks{wyq13889702166@163.com}

\affiliation{
$^1$Department of Physics, Liaoning Normal University, Dalian 116029, China\\
$^2$Center for Theoretical and Experimental High Energy Physics, Liaoning Normal University, Dalian 116029, China
}

\begin{abstract}

New pseudoscalars, axion-like particles (ALPs), provide the exciting target for present and future collider-based experiments. Search for ALPs is performed in this paper via the $W^{+}W^{-}$ fusion process  $e^{-}e^{+}\rightarrow\nu_{e}\overline{\nu_{e}}a\rightarrow\nu_{e}\overline{\nu_{e}}f\overline{f}$ at the $1$ TeV ILC corresponding to an integrated luminosity of $1$ ab$^{-1}$ and the beam polarization
$P(e^{-}$, $e^{+}) = (-80\%$, $+20\%)$. Owing to the good capability of the ILC in performing b-tagging and the sufficiently large branching ratio of the ALP decaying into a pair of b quarks, the decay channel $a\rightarrow{b\overline{b}}$ is mainly concerned. The prospective sensitivities provided by the ILC on the ALP-fermion coupling as low as $1$ TeV$^{-1}$ and $1.75$ TeV$^{-1}$ are derived at $95\%$ confidence level in the ALP mass intervals $37-50$ GeV and $52-300$ GeV, respectively. Our results will help to probe significant parameter space in an unexplored region beyond the existing constraints.

\end{abstract}

\maketitle

\section{Introduction}

An important issue in particle physics at present is the search for new phenomena to address the shortcomings of the greatly successful Standard Model (SM)~\cite{Peccei:1977hh, Peccei:1977ur, Weinberg:1977ma, Wilczek:1977pj,Davidson:1981zd}. Extra spin zero particles, axion-like particles (ALPs), are actually raised by candidate solutions of the main and pressing problems. ALPs provide an interesting link to the puzzle of dark matter~\cite{Preskill:1982cy} and are ubiquitous in the BSM theories such as string theory~\cite{Witten:1984dg} and supersymmetry~\cite{Frere:1983ag}. ALPs originated from a spontaneously broken global symmetry are singlet states under the SM and are capable of communicating with the SM particles. They generally have similar interactions as axions~\cite{Dine:1981rt, Zhitnitsky:1980tq}, but there is no strict relationship between the mass and the coupling. Searching for ALPs has gathered considerable attention in recent years, being the window for new physics beyond the SM (BSM).

Searching for ALPs can be performed using the existing experiments at our disposal or future facilities being built for different reasons. Although cosmological limits are severe \cite{Ringwald:2014vqa, Marsh:2015xka}, they can be avoided for the heavier ALPs subjecting to collider studies. Up to now, most phenomenological ALP analyses have concentrated on their couplings to gluons, photons and massive gauge bosons~\cite{Ghebretinsaea:2022djg, Mimasu:2014nea, Yue:2019gbh, Zhang:2021sio, Yue:2021iiu, Wang:2022ock, Yue:2022ash, Bao:2022onq, Han:2022mzp, Yue:2023mew}. ALPs could also couple to the SM fermions and cause novel signatures at collider experiments. Several interesting bounds on the ALP-fermion interactions already exist. For example, Refs.~\cite{Brivio:2017ije, Esser:2023fdo} investigate the direct probe to the coupling of the light ALP with top quarks via the process $pp\rightarrow{t\overline{t}a}$, in which the ALP is taken as the missing transverse energy. The indirect limits arising from the ALP-top contribution to the loop-induced gluon-gluon fusion process have also been studied~\cite{Bonilla:2021ufe, Esser:2023fdo}. While most of the bounds on the ALP-fermion couplings come from the decays of mesons for the light ALPs~\cite{Bauer:2021mvw,Guerrera:2022ykl}.

The constraints on the couplings of the ALP with fermions are significantly weaker as the ALP mass increases. The process $pp\rightarrow{t\overline{t}a(a\rightarrow{e^{+}e^{-}/\mu^{+}\mu^{-}})}$ has been studied at the LHC~\cite{CMS:2019lwf}. A search for $Z$ decaying into a photon and a resonance ALP with  mass between $60$ and $84$ GeV at the LEP is performed in Ref.~\cite{OPAL:1991acn}, where the ALP subsequently decays into a pair of leptons. These researches to date call for further phenomenological explorations of the fermionic signals associated to the ALP production~\cite{Liu:2022tqn, Calibbi:2022izs}. In this study, we will focus on probing the couplings of ALP with fermions by considering the $W^{+}W^{-}$ fusion process with the decays of the ALP into a pair of fermions $e^{-}e^{+}\rightarrow\nu_{e}\overline{\nu_{e}}a\rightarrow\nu_{e}\overline{\nu_{e}}f\overline{f}$ at the $1$ TeV International Linear Collider (ILC)~\cite{ILC:2019gyn, Bambade:2019fyw} for an integrated luminosity of $1$ ab$^{-1}$ and beam polarization $P(e^{-}$, $e^{+}) = (-80\%$, $+20\%)$, especially targeting for the $a\rightarrow{b\overline{b}}$ decay mode. The clean environment, low backgrounds, adjustable beam energies and polarizations offered by the ILC with the ability to discover the new physics clues will allow such a future  $e^{+}e^{-}$ collider being a powerful complement to the LHC. The potential sensitivities on the parameter space obtained by us at the ILC are better than those given by other future devices. Therefore, our signal process provides an important avenue to search for ALPs at the ILC.

In the remainder of this paper, we first start from the most general effective ALP Lagrangian at dimension-$5$ order and have an overview of the ALP production in our promising channels in Sec. II. Next, the prospects for detecting ALPs at the ILC are presented in Sec. III, as well as the discussions of the validity range of our numerical results comparing with current experimental bounds and other future projections on the ALP-fermion coupling. Finally, the concluding remarks are offered in Sec. IV.

\section{Theoretical setup}

A general Lagrangian for interactions between the pseudoscalar ALP and the SM particles with dimension up to five is~\cite{Georgi:1986df,Brivio:2017ije,Bauer:2017ris}

\begin{eqnarray}
\begin{split}
\mathcal{L}_{\text{eff}}
	=
    &\frac{1}{2} (\partial^\mu a)(\partial_\mu a) - \frac{1}{2} m_a^2 a^2 + \frac{\partial^{\mu} a}{f_a} \sum_{\substack{\psi=Q_L,\,Q_R, \\\,L_L,\,L_R}} \bar\psi \gamma_\mu X_\psi \psi \\
    &- C_{\tilde{W}} \frac{a}{f_a} W^{i}_{\mu\nu} \tilde{W}^{i\mu\nu} - C_{\tilde{B}} \frac{a}{f_a} B_{\mu\nu} \tilde{B}^{\mu\nu}\text{,}
\end{split}
\end{eqnarray}
where $W^{i}_{\mu\nu}$ and $B_{\mu\nu}$ represent the strength tensors associated with the $SU(2)$ and $U(1)$ gauge symmetries of the SM, respectively. The dual field strengths are expressed by $\tilde{X}^{\mu\nu}=\frac{1}{2}\epsilon^{\mu\nu\lambda\kappa}X_{\lambda\kappa}$ $(X = W^{i}$, $B)$. The mass and decay constant of the ALP are, respectively, $m_a$ and $f_a$. The coefficients $C_{\tilde{W}}$ and $C_{\tilde{B}}$ along with $m_a$ and $f_a$ are considered as free parameters. $X_\psi$ are Hermitian matrices in flavour space. In this work, the coupling of ALP with gluons is not considered.

\begin{figure}[H]
\begin{center}
\centering\includegraphics [scale=0.6] {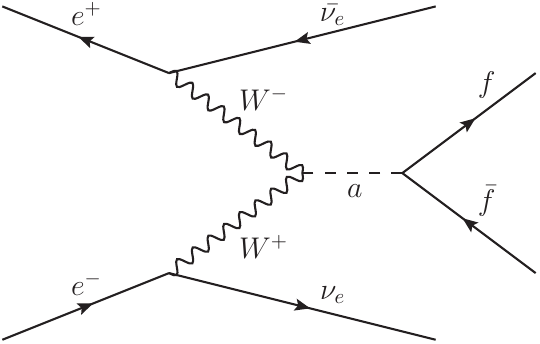}
\caption{The Feynman diagram of the $W^{+}W^{-}$ fusion process $e^{-}e^{+}\rightarrow\nu_{e}\overline{\nu_{e}}a\rightarrow\nu_{e}\overline{\nu_{e}}f\overline{f}$.}
\label{feymadia}
\end{center}
\end{figure}

After electroweak symmetry breaking (EWSB), writing down the following Lagrangian terms related to our study

\begin{eqnarray}
\mathcal{L}_{\text {eff }} &\supset& i g_{a\psi} a \sum_{\psi=Q,\,L}  m^{\text{diag}}_{\psi}\, \bar{\psi} \gamma_5 \psi -\dfrac{1}{4}g_{aWW}aW_{\mu\nu}\tilde{W}^{\mu\nu}\text{,}
\end{eqnarray}
in which ALPs are expected to interact with the SM fermions and $W^{\pm}$ bosons. The
effective ALP-fermion interaction is depicted by the coupling parameter $g_{a\psi}$. The symbol $\psi$ in the first term runs over all the SM fermions other than the neutrinos and $m^{\text{diag}}_{\psi}$ denotes the fermion mass matrix. The coupling parameter $g_{aWW}$ is expressed by the
coefficient $C_{\tilde{W}}$ in Eq.(1) with $f_a$, which is expressed as $g_{aWW} = \frac{4}{f_a} C_{\tilde{W}}$. In this work, we focus on detecting ALPs at the ILC through the $W^{+}W^{-}$ fusion process $e^{-}e^{+}\rightarrow\nu_{e}\overline{\nu_{e}}a$ followed by $a$ decaying to fermions, as shown by the Feynman diagram in FIG.~\ref{feymadia}. For illustrative purposes, the limits on $g_{aWW}$ derived in Ref.~\cite{Carra:2021ycg} describing the search for ALPs at the LHC via the gluon-gluon fusion process are applied in our analysis and the assumption of $C_{\tilde{W}} = C_{\tilde{B}}$ is regarded. Accordingly, the parameter space is spanned by the ALP mass $m_a$ and the coupling parameter $g_{a\psi}$.

\begin{figure}[H]
\begin{center}
\subfigure[]{\includegraphics [scale=0.3] {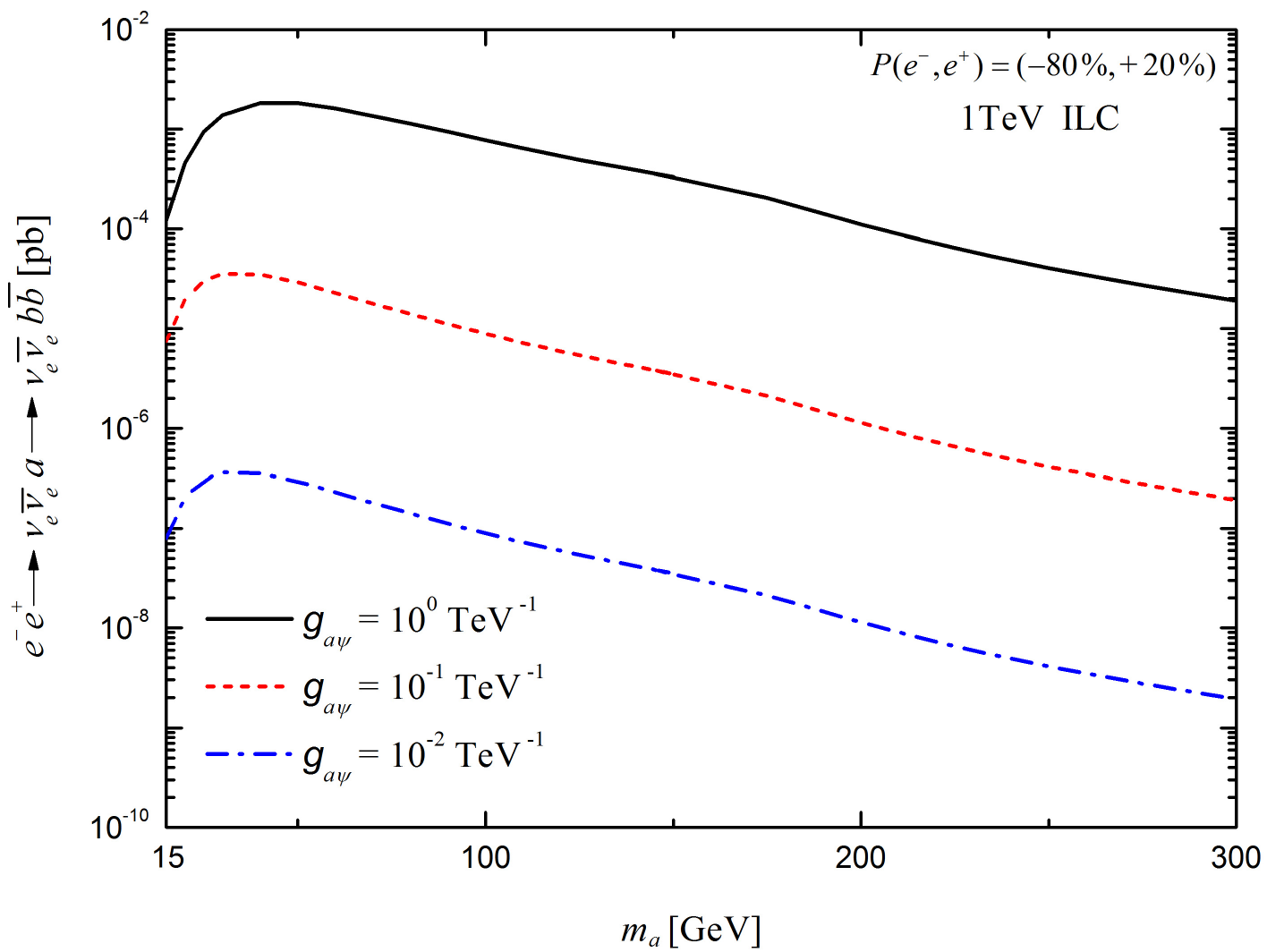}}
\subfigure[]{\includegraphics [scale=0.3] {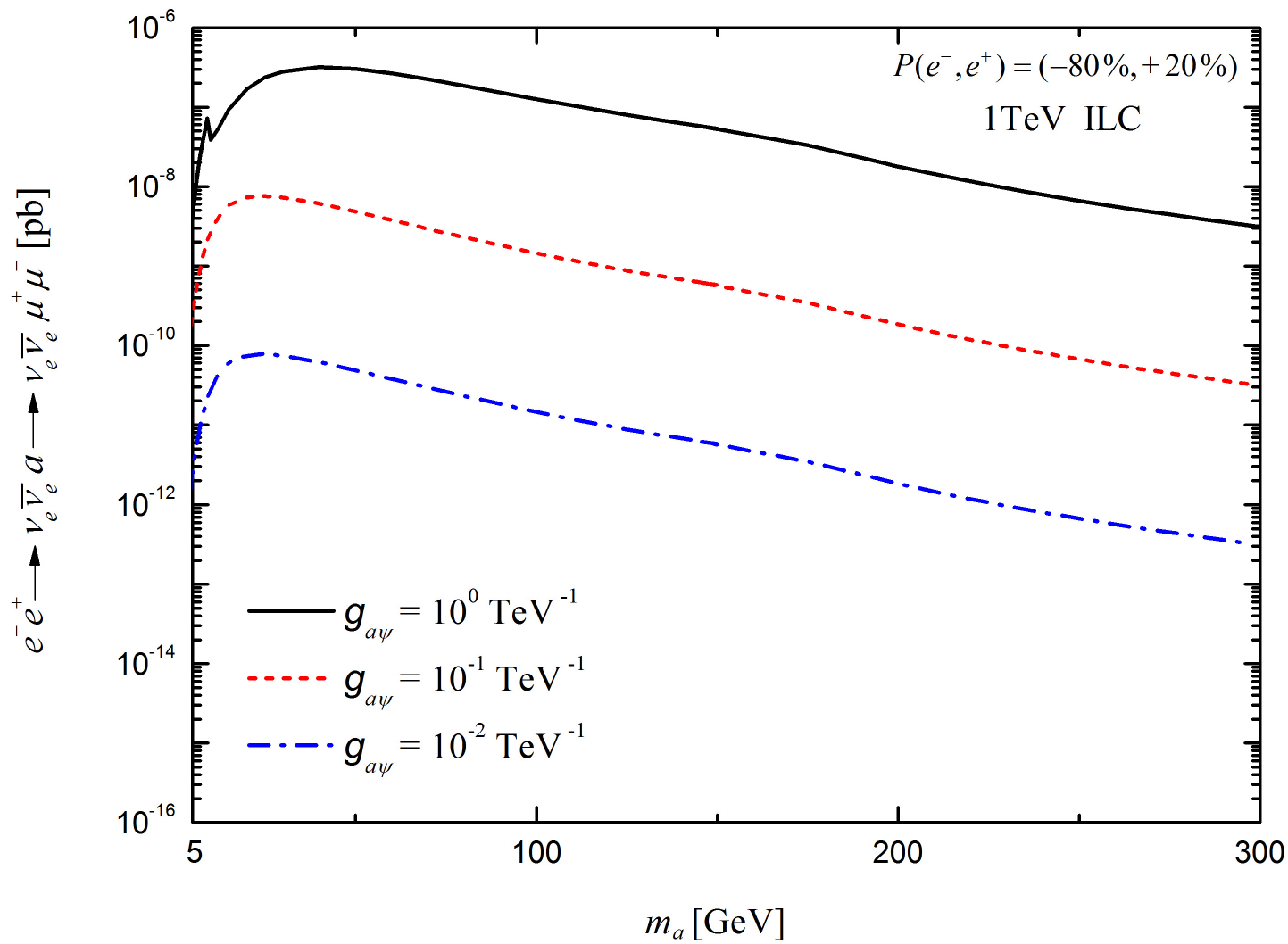}}
\caption{The production cross sections of the $W^{+}W^{-}$ fusion processes $e^{-}e^{+}\rightarrow\nu_{e}\overline{\nu_{e}}a\rightarrow\nu_{e}\overline{\nu_{e}}b\overline{b}$ (a) and $e^{-}e^{+}\rightarrow\nu_{e}\overline{\nu_{e}}a\rightarrow\nu_{e}\overline{\nu_{e}}\mu^{+}\mu^{-}$ (b) as functions of the ALP mass.}
\label{maandxsection}
\end{center}
\end{figure}

For simplicity, the $t\overline{t}$ signal production mode is not considered in our study because of the relatively unclean backgrounds in such case and the ALP with mass below the top pairs threshold is taken into account. The various decay rates for different modes of the ALP decaying into pairs of the SM fermions are shown in FIG. 3 of Ref.~\cite{Bauer:2017ris}. Although it can be seen that decay rates of the $a\rightarrow{c\overline{c}}$ and $a\rightarrow{\tau^{+}\tau^{-}}$ channels are comparatively large, the fact that c-jets are not tagged in the detector as efficiently as b-jets and the tau lepton decays rapidly, making their identifications very challenging. The cross sections of the processes $e^{-}e^{+}\rightarrow\nu_{e}\overline{\nu_{e}}a\rightarrow\nu_{e}\overline{\nu_{e}}f\overline{f}$ with $f = \mu$, $e$, $s$, $d$, $u$ are four to nine orders of magnitude smaller than those of the signal process in which the ALP decays into $b\overline{b}$ pairs for $g_{a\psi} = 1$ TeV$^{-1}$, which is due to the fact that the ALP-fermion coupling is proportional to the fermion mass and therefore larger for the heavier generations. To demonstrate such a gap more dramatically, variation of the cross sections for the $W^{+}W^{-}$ fusion processes $e^{-}e^{+}\rightarrow\nu_{e}\overline{\nu_{e}}a\rightarrow\nu_{e}\overline{\nu_{e}}b\overline{b}$ and $e^{-}e^{+}\rightarrow\nu_{e}\overline{\nu_{e}}a\rightarrow\nu_{e}\overline{\nu_{e}}\mu^{+}\mu^{-}$ with respect to the appropriate mass of the ALP is shown in FIG.~\ref{maandxsection}, in which the cross sections of the former are obviously much larger than those of the latter when $g_{a\psi}$ is taken to be the same value.

Therefore, the process of $e^{-}e^{+}\rightarrow\nu_{e}\overline{\nu_{e}}a\rightarrow\nu_{e}\overline{\nu_{e}}b\overline{b}$ is mainly studied in our work and the ALP with mass ranging from $15$ to $300$ GeV is desirable. The ILC with b-tagging capability of an unprecedented excellence laying out the center of mass energy upgraded to $1$ TeV with an integrated luminosity of $1$ ab$^{-1}$ is taken in our analysis. Polarization is an essential ingredient in the experimental program and polarized beams with $P(e^{-}$, $e^{+}) = (-80\%$, $+20\%)$ will be employed, under which the cross sections of the signal process are relatively large. The simulation is performed with \verb"MadGraph5_aMC@NLO"~\cite{Alwall:2014hca}, which generates events for the production of ALPs at the $e^{+}e^{-}$ collider. The numerical results above have been imposed with the basic cuts described in the next section.

\section{The probe of ALPs at the ILC}

Both the signal and backgrounds need to pass by the following basic cuts (preselection cuts) initially in our simulation. The transverse momentum $p_{T}$ of b-jets needs to be larger than $5$ GeV, and that of light flavor jets requires greater than $20$ GeV. The cone size $\Delta R$ as well as the absolute value of the pseudorapidity $\left| \eta \right|$ for both b-jets and light flavor jets are larger than $0.4$ and less than $5$, respectively.

The $b\overline{b}$ pairs together with missing energy can be shown as the final states for the signal process. The corresponding backgrounds mainly include $e^{-}e^{+}\rightarrow\nu_{l}\overline{\nu_{l}}b\overline{b}$ and $e^{-}e^{+}\rightarrow\nu_{l}\overline{\nu_{l}}jj$. The parton shower generator \verb"PYTHIA"~\cite{Sjostrand:2014zea} and the fast detector simulator \verb"DELPHES"~\cite{deFavereau:2013fsa} are also used to do the simulation below. The kinematic and cut-based analysis is realized by \verb"MadAnalysis5"~\cite{Conte:2012fm, Conte:2014zja, Conte:2018vmg}. We find that the coupling parameter $g_{a\psi}$ has a certain influence on the distributions of observables in the final states for the lighter $m_a$. The larger the $g_{a\psi}$ is, the wider the distributions will be at the same mass benchmark point of the signal, especially in the case of the ALP mass less than $50$ GeV. This effect can be ignored with the increasing ALP mass. Hence the mass range of the ALP under consideration is split into two intervals for studying, taking $m_a = 50$ GeV as a breakpoint.

\begin{figure}[H]
\begin{center}
\subfigure[]{\includegraphics [scale=0.5] {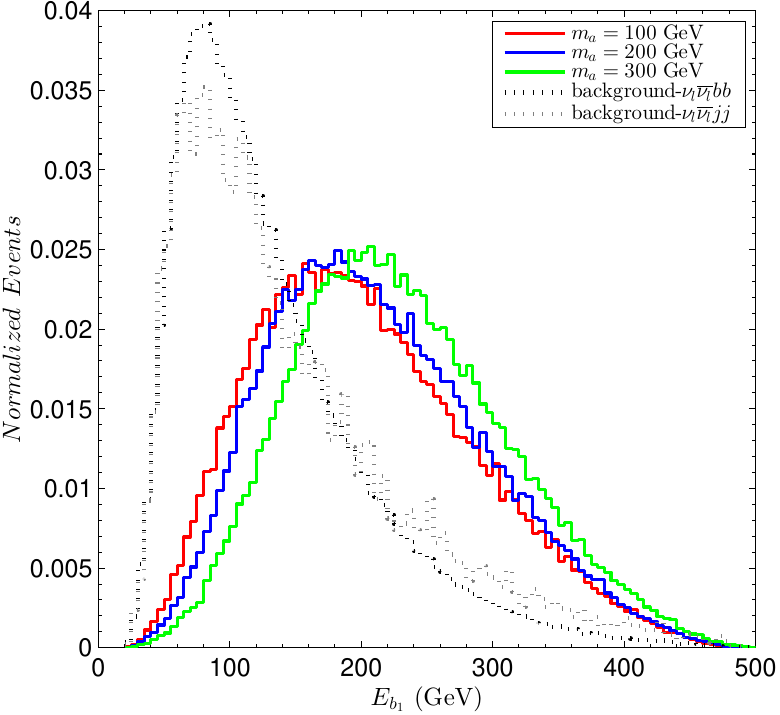}}
\hspace{0.2in}
\subfigure[]{\includegraphics [scale=0.5] {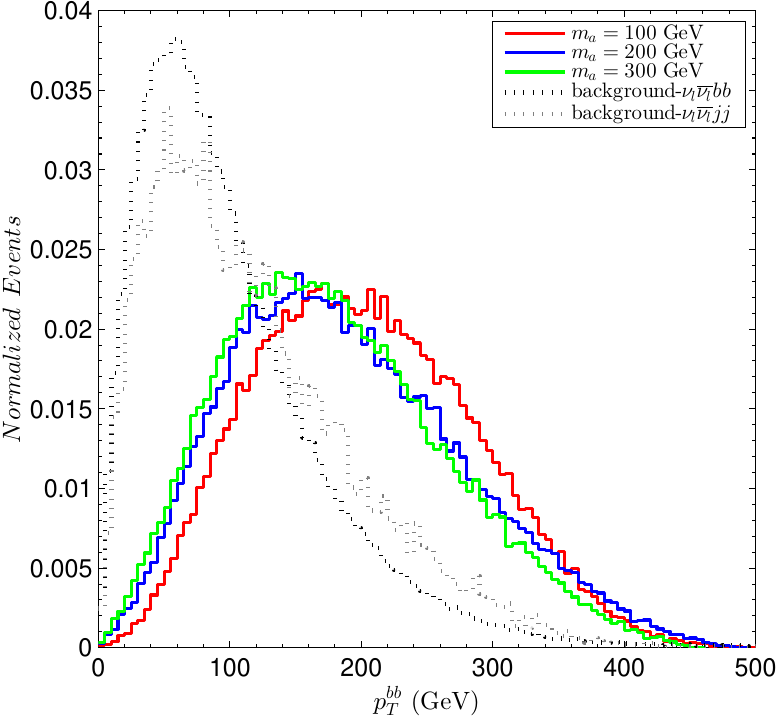}}
\caption{The distributions of $E_{b_{1}}$ (a) and $p_{T}^{bb}$ (b) in the ALP signal and the SM background events for $m_a = 100$, $200$ and $300$ GeV at the $1$ TeV ILC with $\mathcal{L}=$ $1$ ab$^{-1}$ and $P(e^{-}$, $e^{+}) = (-80\%$, $+20\%)$.}
\label{distribution}
\end{center}
\end{figure}
\begin{table}[H]
\begin{center}
\caption{The optimized selection cuts for the signal and backgrounds.}
\label{table:bbcut}
\begin{tabular}
[c]{c|c c c c c}\hline \hline
\multirow{2}{*}{Cuts}    & \multicolumn{2}{c}{ Mass }   \\
\cline{2-3}
	       &~~~~~~~$15$ GeV $\leq$ $m_a\leq50$ GeV~~~~~~~      &~~~~~~~$50$ GeV $<$ $m_a\leq300$ GeV~~~~~~~      \\ \hline
	Cut 1      &  $N_b\geq2$  & $N_b\geq2$ \\

	Cut 2         &  $m_{bb}<80$ GeV   & $E_{b_{1}}>120$ GeV  \\

	Cut 3         &  $\Delta R _{bb}<1$  & $p_{T}^{bb}>100$ GeV    \\

    Cut 4         &  $\slashed E_{T}>40$ GeV & $-$    \\

	Cut 5         &  $\Delta \theta_{bb}<0.8$  & $-$   \\ \hline \hline

\end{tabular}
\end{center}
\end{table}

\begin{table}[H]
\begin{center}
\caption{The cross sections for the signal and backgrounds at benchmark points after the step-by-step optimized selection cuts employed at the $1$ TeV ILC with $P(e^{-}$, $e^{+}) = (-80\%$, $+20\%)$ for $g_{a\psi} = 10$ TeV$^{-1}$. The statistical significance $SS$ is computed for the integrated luminosity of $1$ ab$^{-1}$.}
\label{table:ss}
\begin{tabular}
[c]{c|c c c}\hline \hline
\multirow{2}{*}{Cuts}    & \multicolumn{3}{c}{Cross sections for signal (bg-$\nu_{l}\overline{\nu_{l}}b\overline{b}$, bg-$\nu_{l}\overline{\nu_{l}}jj$) [pb]}   \\
\cline{2-4}
	       &~~~$m_a=100$ GeV&~~~~~~$m_a=200$ GeV&~~~~~$m_a=300$ GeV    \\ \hline
	Basic Cuts   & ~~ \makecell{$5.4310\times10^{-3}$\\$(0.6889$, $0.8857)$}  & ~~~~~ \makecell{$3.1891\times10^{-3}$\\$(0.6889$, $0.8857)$} & ~~~~~\makecell{$1.0866\times10^{-3}$\\$(0.6889$, $0.8857)$} \\

	Cut 1        & ~~ \makecell{$1.6686\times10^{-3}$\\$(0.1809$, $4.3155\times10^{-3})$}  & ~~~~ \makecell{$1.3018\times10^{-3}$\\$(0.1809$, $4.3155\times10^{-3})$} & ~~~~~\makecell{$4.9840\times10^{-4}$\\$(0.1809$, $4.3155\times10^{-3})$}  \\

	Cut 2         & ~~ \makecell{$1.3973\times10^{-3}$\\$(0.0835$, $2.1934\times10^{-3})$}  & ~~~~ \makecell{$1.1422\times10^{-3}$\\$(0.0835$, $2.1934\times10^{-3})$} & ~~~~~\makecell{$4.6240\times10^{-4}$\\$(0.0835$, $2.1934\times10^{-3})$}    \\

    Cut 3         & ~~ \makecell{$1.3170\times10^{-3}$\\$(0.0535$, $1.6275\times10^{-3})$}  & ~~~~ \makecell{$1.0191\times10^{-3}$\\$(0.0535$, $1.6275\times10^{-3})$} & ~~~~~\makecell{$3.9370\times10^{-4}$\\$(0.0535$, $1.6275\times10^{-3})$}    \\  \hline

    $SS$  & ~~$5.545$ & ~~~~$4.300$ & ~~~~~$1.671$ \\ \hline \hline

\end{tabular}
\end{center}
\end{table}
Some variables are targeted to distinguish the signal from the overwhelming backgrounds in the mass range of ALP from $15$ to $50$ GeV, such as the invariant mass $m_{bb}$, the cone size $\Delta R _{bb}$, the missing transverse energy $\slashed E_{T}$ and the angular separation $\Delta \theta _{bb}$. Although the $m_{bb}$ distribution peak is mostly around the ALP mass for the signal process, the corresponding regions are almost barely covered by the backgrounds. The light ALP always decays to the collimated $b\overline{b}$ pairs, while the distributions of the angular separation between $b\overline{b}$ pairs in the backgrounds go wide. Powerful observables of the energy of the hardest b-jet $E_{b_{1}}$ and the transverse momentum of the reconstructed ALP $p_{T}^{bb}$ are useful to pick out the signal in the interval between $50$ and $300$ GeV of the ALP mass, kinematic distributions of which are displayed in FIG.~\ref{distribution} for several $m_a$ benchmark points in both the signal as well as the dominant SM backgrounds among the set of simulated MC samples. It can be seen that the energy from the hardest b-jet in the majority of the background events is softer than that of the hardest b-jet in the signal case.

The optimized selection cuts for the identification of the signal and suppression of the backgrounds are listed in TABLE~\ref{table:bbcut} based on the above analysis with the number cuts being taken as the first step filter. In TABLE~\ref{table:ss}, we summarize the cross sections of the signal and backgrounds for several representative ALP mass benchmark points with $g_{a\psi} = 10$ TeV$^{-1}$ after the improved cuts applied at the $1$ TeV ILC, in which the $\nu_{l}\overline{\nu_{l}}b\overline{b}$ background and the $\nu_{l}\overline{\nu_{l}}jj$ background are labeled as ``bg-$\nu_{l}\overline{\nu_{l}}b\overline{b}$" and ``bg-$\nu_{l}\overline{\nu_{l}}jj$", respectively. The statistical significance $SS$ calculated by the formula $SS=N_{sig}/\sqrt{N_{sig} + N_{bg}}$ for an integrated luminosity of $1$ ab$^{-1}$ is given in the last row of TABLE~\ref{table:ss}, where $N_{sig}$ and $N_{bg}$ refer to the number of signal and background events, separately. The value of $SS$ can achieve $5.545$ when $m_a = 100$ GeV.

\begin{figure}[H]
\begin{center}
\centering\includegraphics [scale=0.5] {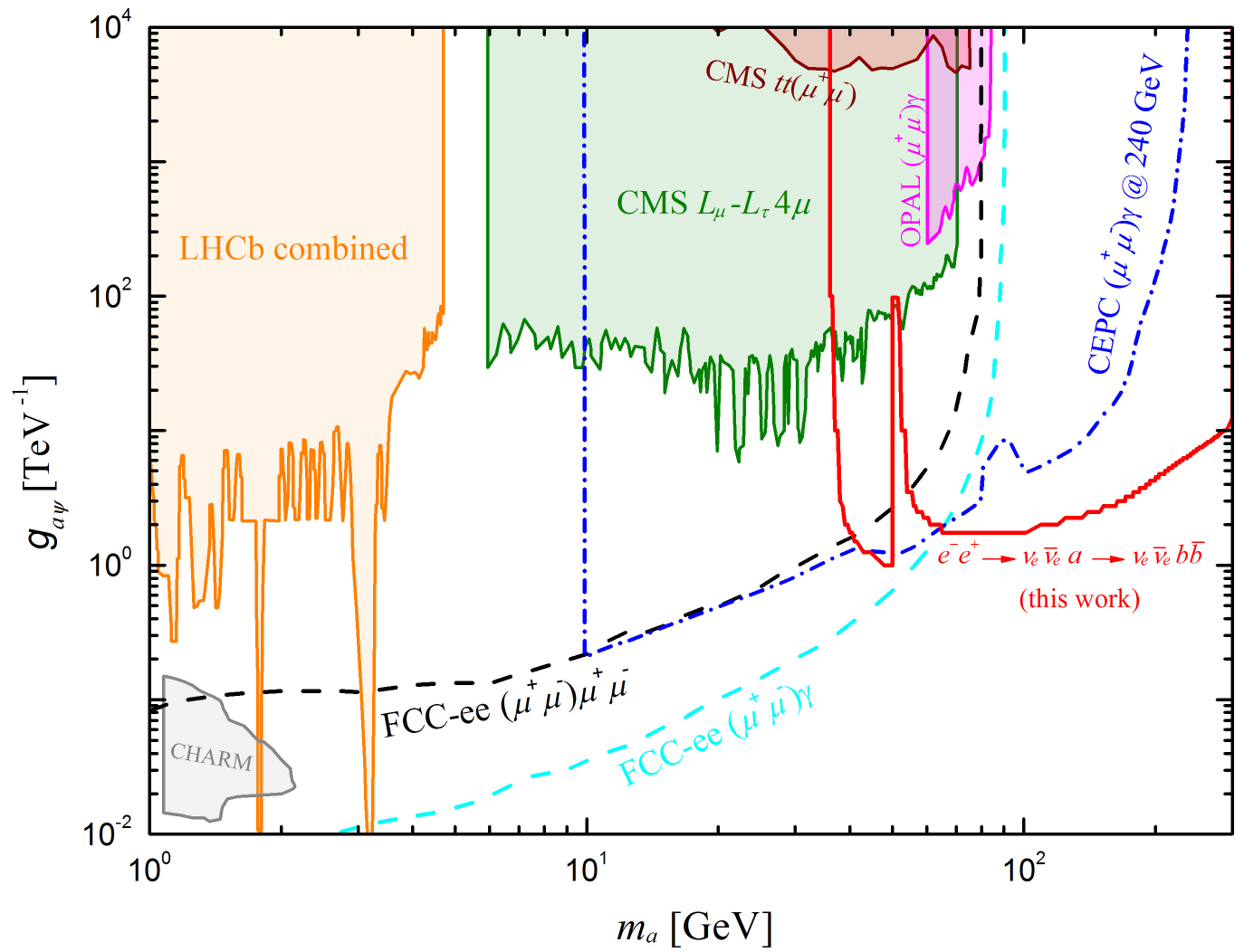}
\caption{Constraints on the ALP-fermion couplings derived from the past collider experiments at $95\%$ C.L. together with the sensitivities of proposed searches in the ALP parameter space, in which the reaches of our results are indicated by the red line.}
\label{ma-gapsi-plane}
\end{center}
\end{figure}

The projected sensitivity reaches for the signal process $e^{-}e^{+}\rightarrow\nu_{e}\overline{\nu_{e}}a\rightarrow\nu_{e}\overline{\nu_{e}}b\overline{b}$ at the $1$ TeV ILC with $1$ ab$^{-1}$ and $P(e^{-}$, $e^{+}) = (-80\%$, $+20\%)$ are shown in FIG.~\ref{ma-gapsi-plane} by the red line. The ALP-fermion couplings can be constrained by several collider experiments, also displayed in the colored shaded regions in the $m_a-g_{a\psi}$ plane of FIG.~\ref{ma-gapsi-plane}, in which the promising sensitivities of other future $e^{+}e^{-}$ colliders to the couplings of the ALP with fermions are described by dashed lines. Our results, complementing the researches for the ALP-fermion interactions by the CMS and OPAL, demonstrate that the promising sensitivities to the couplings of ALP with fermions can be improved to $1$ TeV$^{-1}$ for the ALP in the mass region of $37$ to $50$ GeV as well as down to $1.75$ TeV$^{-1}$ for $m_a$ between $52$ and $300$ GeV. Compared to the projected future sensitivities of other experiments, our signal process is more sensitive to probe the ALP with mass $65-300$ GeV.

The powerful collider constraints for the ALP mass below $5$ GeV come from the CHARM (grey region)~\cite{Dobrich:2018jyi} and the LHCb (orange region)~\cite{LHCb:2015nkv, LHCb:2016awg} experiments through decays of heavy mesons. The bounds from the CMS detector at the LHC (dark red and olive green regions)~\cite{CMS:2018yxg, CMS:2019lwf} and the OPAL detector at the LEP (magenta region)~\cite{OPAL:1991acn} together rule out a portion of the parameter space. The CMS collaboration has executed the search for the $4\mu$ final states on $U(1)_{L_{\mu}-L_{\tau}}$~\cite{CMS:2018yxg} as well as given the limits on the process of generating a light scalar or pseudoscalar in association with a pair of top quarks~\cite{CMS:2019lwf}, resulting in a significant increase in the excluded area, while the OPAL search for on-shell $Z\rightarrow(\mu^{+}\mu^{-})\gamma$ plays a supplementary role~\cite{OPAL:1991acn}. The CEPC and FCC-ee, regarded as the future Tera-Z and Higgs factories, will be able to detect interesting regions of the uncovered parameter space via the probes with final states $(\mu^{+}\mu^{-})\gamma$ and $(\mu^{+}\mu^{-})\mu^{+}\mu^{-}$ (dashed lines)~\cite{Liu:2022tqn}. In a nutshell, the sensitivity limits derived from our work would nicely complement and further broaden the already physics programs in the cases of $m_a \in [37$, $50]$ GeV and $[52$, $300]$ GeV.

\section{Conclusions}

Many studies have been conducted to search for new physics over the past few years. Theoretically and phenomenologically, the ALP represents an attractive possibility with good motivation and may well show up first at colliders. A broader range of detectable parameter space on the ALP-fermion couplings can be realized at the ILC with highly polarized beams and low background levels.

In this paper, we have explored the search for the ALP via the process $e^{-}e^{+}\rightarrow\nu_{e}\overline{\nu_{e}}a\rightarrow\nu_{e}\overline{\nu_{e}}f\overline{f}$ at the $1$ TeV ILC with $1$ ab$^{-1}$ and $P(e^{-}$, $e^{+}) = (-80\%$, $+20\%)$ beam polarization, in particular dedicated to the decay channel $a\rightarrow{b\overline{b}}$. The results we obtained show that a part of the currently unconstrained parameter space can be covered by the proposed ILC, for which the corresponding reaches are $1$ TeV$^{-1}$ and $1.75$ TeV$^{-1}$ from $37$ GeV $\leq$ $m_a\leq50$ GeV and $52$ GeV $\leq$ $m_a\leq300$ GeV, respectively. Our estimates of the sensitivity limits are highly complementary to the LHC existing bounds and provide valuable insights for future experiments aimed at detecting ALPs.

\section*{ACKNOWLEDGMENT}

This work was partially supported by the National Natural Science Foundation of China under Grants No. 11875157 and No. 12147214.


\bibliography{ALPref}

\end{document}